\pdfoutput=1
\documentclass{llncs}

\usepackage{amsfonts,amsmath}

\usepackage{graphicx}
\usepackage{mathptmx}
\usepackage{tikz-qtree}
\usepackage{pgfplots}
\usepackage{filecontents}
\usepackage{csvsimple}
\usepackage{amsfonts,amsmath}
\usepackage{url}
\usepackage{verbatim}
\usepackage{color}

\definecolor{lgray}{gray}{0.95}
\definecolor{lblue}{rgb}{0.90,0.90,1.00}
\definecolor{lyellow}{rgb}{1.00,1.00,0.70}

\usepackage{listings}
\newtheorem{ex}{Example}

\newenvironment{codex}{\small\verbatim}{\endverbatim\normalsize}

\lstnewenvironment{code}
    {\lstset{}%
      \csname lst@SetFirstLabel\endcsname}
    {\csname lst@SaveFirstLabel\endcsname}
\newcommand{\BI}[0]{\begin{itemize}}
\newcommand{\EI}[0]{\end{itemize}}
\newcommand{\I}[0]{\item}
\newcommand{\BE}[0]{\begin{enumerate}}
\newcommand{\EE}[0]{\end{enumerate}}

\newcommand{\BX}[0]{\begin{ex}}
\newcommand{\EX}[0]{\end{ex}}
\newcommand{\BV}[0]{\begin{verbatim}}
\newcommand{\EV}[0]{\end{verbatim}}

\newcommand{\BF}[0]{\begin{filecontents*}{data.csv}}

\newcommand{\BQ}[0]{\color{blue}\begin{quote}}
\newcommand{\EQ}[0]{\end{quote}\color{black}}

\def \bscale1 {0.25}
\def \bscale {0.25}

\newcommand{\imp}{\rightarrow}
\newcommand{\Imp}{\Rightarrow}

\newcommand{\bk}{\ensuremath{\mathbf{K}}}

\begin{document}

\title{Modality Definition Synthesis for Epistemic Intuitionistic Logic via a Theorem Prover}
\author{Paul Tarau}

\institute{
   {Department of Computer Science and Engineering}\\
   {University of North Texas}\\
   {\em paul.tarau@unt.edu}
}

\maketitle

\begin{abstract}
We derive a Prolog theorem prover for an
Intuitionistic Epistemic Logic by
starting from the sequent calculus {\bf G4IP}
that we extend with operator definitions 
providing an embedding  in 
intuitionistic propositional logic ({\bf IPC}).

With help of a candidate definition formula generator, we discover
 epistemic operators for which  axioms and theorems of 
Artemov and Protopopescu's {\em Intuitionistic Epistemic Logic} ({\bf IEL})
hold and formulas  expected to be non-theorems fail.

We compare the embedding of {\bf IEL} in {\bf IPC} with a similarly discovered successful
embedding of Dosen's double negation modality, judged inadequate
as an epistemic operator.
Finally, we discuss the
failure of the {\em necessitation rule} for an otherwise
successful {\bf S4} embedding and share our thoughts
about the intuitions explaining
these differences between epistemic and alethic modalities
in the context of the Brouwer-Heyting-Kolmogorov semantics
of intuitionistic reasoning and knowledge acquisition.
\end{abstract}


\section{Introduction}

Epistemic Logic Systems have been derived often in parallel and sometime as afterthoughts of alethic Modal Logic Systems, in which modalities are defined by axioms and additional inference rules extending classical logic.

With the advent of Answer Set Programming ({\bf ASP}) epistemic logics hosted in this framework like e.g., \cite{gelfond_intro,baral10,shen16} show that   {\em intermediate logics
\footnote{Logics stronger than intuitionistic but weaker than classical.}} 
can express epistemic operators  by extending the underlying logic 
with definition of epistemic operators.

Steps\footnote{Actually infinitely many, as there's an infinite lattice of intermediate
logics between classical and intuitionistic logic.}
further below classical logic or {\bf ASP} are taken  in recent work \cite{iel16},
based on the
the Brouwer-Heyting-Kolmogorov  ({\bf BHK}) 
 view of intuitionistic logic  that takes 
into account  the constructive nature of knowledge, 
modeling more accurately the connection between proof systems and
the related mental processes.
 
Along these lines, our inquiry into epistemic logic will focus on knowledge vs. truth seen as intuitionistic provability.

Like in the case of embedding epistemic operators into {\bf ASP} systems, but with a ``machine-learning'' twist, we will design a synthesis mechanism for epistemic operators via embedding in {\bf IPC}. For this purpose we will generate candidate formulas that verify axioms, theorems and rules and fail on expected non-theorems.
We will also show that this   view generalizes to a mechanism for
discovering the right formalization of a given modal logic.

Our starting point is Artemov and Protopopescu's {\em Intuitionistic Epistemic Logic} ({\bf IEL})
\cite{iel16} that will provide the axioms, theorems and non-theorems stating the requirements
that must hold for the definitions extending {\bf IPC}.
The discovery mechanism will also bring up Dosen's interpretation of double negation 
\cite{dosen_dneg} as a potential epistemic operator and we will look  into applying the same discovery mechanisms to find an embedding of modal logic {\bf S4} in {\bf IPC}, with special focus on the impact of the {\em necessitation rule},
which requires that all theorems of the logic are necessarily true.

\subsubsection*{The rest of the paper is organized as follows.}

Section \ref{iel} overviews  Artemov and Protopopescu's {\em Intuitionistic Epistemic Logic} ({\bf IEL}).
Section \ref{prover} introduces the {\bf G4IP} sequent calculus prover for Intuitionistic Propositional Logic ({\bf IPC}).
Section \ref{gen} describes the generator for candidate formulas extending 
{\bf IPC} with modal operator definitions.
Section \ref{embeil} explains the discovering of the definitions that ensure the embedding of {\bf IEL} into {\bf IPC} and 
the discovering of the embedding of Dosen's 
double negation as a modality operator.
It also  discusses the intuitions behind the embedding of  {\bf IEL}, including the epistemic equivalent of the necessity rule in {\bf IPC} and the adequacy of this embedding as a constructive mechanism for reasoning about knowledge.
Section \ref{s4} studies the case of the {\bf S4} modal logic and the failure of the necessity rule, indicating the difficulty of embedding it in {\bf IPC} by contrast to {\bf IEL}.
Section \ref{rel} overviews  some related work and
section \ref{concl} concludes the paper.

The paper is written as a literate SWI-Prolog program with its  extracted code at {\small
 \url{https://raw.githubusercontent.com/ptarau/TypesAndProofs/master/ieltp.pro}}.

\section{Overview of Artemov and Protopopescu's {\bf IEL} logic}\label{iel}

In \cite{iel16} a system for  Intuitionistic Epistemic Logic is introduced that
\begin{quote}
``maintains the original Brouwer-Heyting-Kolmogorov semantics for intuitionism and is consistent with the well-known approach that intuitionistic knowledge be regarded as the result of verification''.
\end{quote}
Instead of the classic, alethic-modalities inspired \bk ~operator for which
\[
\bk A \imp  A\label{CT}
\]
Artemov and Protopopescu argue that{ {\em co-reflection} expresses better the idea of {\em constructivity of truth}
\[
A \imp \bk A\label{CT}
\]
They also argue that this applies to both belief and knowledge i.e., that
\begin{quote}
``The verification-based approach allows that justifications more general than proof can be adequate for belief and knowledge''.
\end{quote}

On the other hand, they consider {\em intuitionistic reflection} acceptable, expressing the fact
that ``known propositions cannot be false'':
\[ 
\bk A \imp \neg\neg A
\]

Thus, they position intuitionistic knowledge of $A$ between $A$ and $\neg\neg A$:
\[ A\imp \bk A\imp \neg\neg A \] and given that (via Glivenko's transformation \cite{gliv})
applying double negation to a formula embeds classical propositional calculus
 into {\bf IPC}, they express this view as:
\begin{center}
      \emph{Intuitionistic Truth $\ \ \Imp\ $ Intuitionistic Knowledge $\ \ \Imp\ $ Classical Truth}.
\end{center}

\noindent They axiomatize the system {\bf IEL} as follows.\\


{\bf 1}. Axioms of propositional intuitionistic logic;

{\bf 2}. $\bk(A \imp B) \imp (\bk A \imp \bk B)$; \hfill (distribution)

{\bf 3}. $A \imp \bk A$. \hfill (co-reflection)

{\bf 4}. $\bk A \imp \neg \neg A$ \hfill (intuitionistic reflection)\\

\textbf{Rule} {\em Modus Ponens}.

\medskip

They also argue that a weaker logic of belief (${\bf IEL}^{-}$) is expressed
by considering only axioms {\bf 1,2,3}.


\section{The {\bf G4ip} prover for {\bf IPC}}\label{prover}

We will describe next our lightweight propositional 
intuitionistic theorem prover, that will be used to discover
an embedding of {\bf IEL} into {\bf IPC}.

\subsection{The LJT/G4ip calculus, (restricted here to the implicational fragment)}

Motivated by problems related to loop avoidance in implementing  Gentzen's {\bf LJ} calculus,
Roy Dyckhoff \cite{dy1}  introduces
the following rules for the {\bf G4ip} calculus\footnote{Originally called the {\tt LJT} calculus in \cite{dy1}. Restricted here to its key  implicational fragment.}.\\\\
{\large
\noindent
\begin{math}
LJT_1:~~~~\frac{~}{A,\Gamma ~\vdash~ A}\\\\\\
LJT_2:~~~~\frac{A,\Gamma ~\vdash~ B}{\Gamma ~\vdash~ A\rightarrow B}\\\\\\
LJT_3:~~~~\frac{B,A,\Gamma ~\vdash~ G}{A \rightarrow B,A,\Gamma ~\vdash~ G}\\\\\\ 
LJT_4:~~~~\frac{D \rightarrow B,\Gamma ~\vdash~ C \rightarrow D ~~~~ B,\Gamma ~\vdash~ G}
{ \left( C \rightarrow D \right) \rightarrow B,\Gamma ~\vdash~ G }\\
\end{math}
}\\
Note that $LJT_4$ ensures termination as formulas in the sequent become smaller.
The rules work with the context $\Gamma$
being either a multiset or a set.\\

For supporting negation, one also needs to add $LJT_5$ that deals with the special term $false$. 
Then negation of $A$ is defined as $A \rightarrow false$.\\\\

{\Large
\noindent
\begin{math}
LJT_5:~~~~\frac{~}{false,\Gamma ~\vdash~ G}\\\\\\
\end{math}
}
Rules for conjunction, disjunction and bi-conditional (not shown here) are also part of the calculus.

As it is not unusual with logic formalisms,
the same calculus had been discovered independently in the 50's by Vorob'ev and
in the 80's-90's by Hudelmaier \cite{hud88,hud93}.

\subsection{A Lightweight Theorem Prover for Intuitionistic Propositional Logic}

Starting from the sequent calculus for the  intuitionistic propositional logic in G4ip \cite{dy1}, to
which we have also added rules for the ``\verb~<->~'' relation, we obtain the following lightweight {\bf IPC} prover.
\begin{code}
:- op(525,  fy,  ~ ).
:- op(550, xfy,  & ).    % right associative
:- op(575, xfy,  v ).    % right associative
:- op(600, xfx,  <-> ).  % non associative
\end{code}

\begin{code}
prove_in_ipc(T):-  prove_in_ipc(T,[]).

prove_in_ipc(A,Vs):-memberchk(A,Vs),!.
prove_in_ipc(_,Vs):-memberchk(false,Vs),!.
prove_in_ipc(A<->B,Vs):-!,prove_in_ipc(B,[A|Vs]),prove_in_ipc(A,[B|Vs]).
prove_in_ipc((A->B),Vs):-!,prove_in_ipc(B,[A|Vs]).
prove_in_ipc(A & B,Vs):-!,prove_in_ipc(A,Vs),prove_in_ipc(B,Vs).
prove_in_ipc(G,Vs1):- % atomic or disj or false
  select(Red,Vs1,Vs2),
  prove_in_ipc_reduce(Red,G,Vs2,Vs3),
  !,
  prove_in_ipc(G,Vs3).
prove_in_ipc(A v B, Vs):-(prove_in_ipc(A,Vs);prove_in_ipc(B,Vs)),!.
\end{code}
\begin{code}  
prove_in_ipc_reduce((A->B),_,Vs1,Vs2):-!,prove_in_ipc_imp(A,B,Vs1,Vs2).
prove_in_ipc_reduce((A & B),_,Vs,[A,B|Vs]):-!.
prove_in_ipc_reduce((A<->B),_,Vs,[(A->B),(B->A)|Vs]):-!.
prove_in_ipc_reduce((A v B),G,Vs,[B|Vs]):-prove_in_ipc(G,[A|Vs]).
\end{code}
\begin{code}  
prove_in_ipc_imp((C->D),B,Vs,[B|Vs]):-!,prove_in_ipc((C->D),[(D->B)|Vs]).
prove_in_ipc_imp((C & D),B,Vs,[(C->(D->B))|Vs]):-!.
prove_in_ipc_imp((C v D),B,Vs,[(C->B),(D->B)|Vs]):-!.
prove_in_ipc_imp((C<->D),B,Vs,[((C->D)->((D->C)->B))|Vs]):-!.
prove_in_ipc_imp(A,B,Vs,[B|Vs]):-memberchk(A,Vs).  
\end{code}
 We
validate it first by testing it on the implicational subset, derived via the Curry-Howard isomorphism \cite{padl19}, then against Roy Dyckhoff's Prolog implementation\footnote{
\url{https://github.com/ptarau/TypesAndProofs/blob/master/third_party/dyckhoff_orig.pro}
},
working on formulas  up to size 12. Finally we run it
on human-made tests\footnote{at \url{http://iltp.de}}, on which we get no errors, 
solving correctly 161 problems, with a 60 seconds timeout, compared with the
175 problems solved by Roy Dyckhoff's more refined, 
heuristics-based 400 lines prover,
with the same timeout\footnote{
\url{https://github.com/ptarau/TypesAndProofs/blob/master/tester.pro}
}.
We refer to \cite{padl19} for the derivation steps of variants of this prover
working on the implicational and nested Horn clause fragments of {\bf IPC}.
While more sophisticated tableau-based provers are available for {\bf IPC}
among which we mention the excellent Prolog-based {\tt fCube} 
\cite{fcube}, our prover's compact size and adequate performance will suffice,
given also the space constraints imposed by  literate programming 
nature of this paper.

\section{The definition formula generator}\label{gen}

We start with a candidate formula generator that we will
constrain further to be used for generating 
candidate definitions of our modal operators.

\subsection{Generating Operator Trees}

We  generate all formulas of a given size
by decreasing the available size parameter at each step
when nodes are added to a tree representation of a formula.
Prolog's {\bf DCG} mechanism is used to collect the leaves
of the tree.

\begin{code}
genOperatorTree(N,Ops,Tree,Leaves):-
  genOperatorTree(Ops,Tree,N,0,Leaves,[]).
    
genOperatorTree(_,V,N,N)-->[V].
genOperatorTree(Ops,OpAB,SN1,N3)-->
  { SN1>0,N1 is SN1-1,
    member(Op,Ops),make_oper2(Op,A,B,OpAB)
  },
  genOperatorTree(Ops,A,N1,N2),
  genOperatorTree(Ops,B,N2,N3).
  
make_oper2(Op,A,B,OpAB):-functor(OpAB,Op,2),arg(1,OpAB,A),arg(2,OpAB,B).
\end{code}

\subsection{Synthesizing the definitions of modal operators}

As we design a generic definition discovery mechanism,
we will denote generically our modal operators as follows.
\BI
\I ``\verb~#~'' for  ``$\Box$''=necessary and ``{\bf K}''=known
\I ``\verb~*~'' for ``$\Diamond$''=possible and ``{\bf M}''=knowable
\EI

\noindent After the operator definitions
\begin{code}
:- op( 500,  fy, #).   
:- op( 500,  fy, *).
\end{code}
we specify our generator as covering the usual binary operators
and we constrain it to have at least one of the leaves of its
generated trees to be a variable. Besides the {\tt false} constant
used in the definition of negation, we introduce also
a new constant symbol ``\verb~?~'' assumed not to occur in the language.
Its role will be left unspecified until the possible
synthesized definitions will be filtered. We will
constrain candidate definitions to ensure that axioms and 
selected theorems hold and selected  non-theorems fail.
\begin{code}
genDef(M,Def):-genDef(M,[(->),(&),(v)],[false,?],Def).

genDef(M,Ops,Cs,(#(X):-T)):-
  between(0,M,N),
  genOperatorTree(N,Ops,T,Vs),
  pick_leaves(Vs,[X|Cs]),
  term_variables(Vs,[X]).
\end{code}
Leaves of the generated trees will be picked from a given set.
\begin{code}   
pick_leaves([],_).
pick_leaves([V|Vs],Ls):-member(V,Ls),pick_leaves(Vs,Ls).
\end{code}
We first expand our operator definitions for the ``\verb|~|'' negation and ``\verb~*~''
 modal operator
while keeping atomic variables and the special constant {\tt false} untouched.
\begin{code}
expand_defs(_,false,R) :-!,R=false.
expand_defs(_,A,R) :-atomic(A),!,R= A.
expand_defs(D,~(A),(B->false)) :-!,expand_defs(D,A,B).
expand_defs(D,*(A),R):-!,expand_defs(D,~ (# (~(A))),R).
\end{code}
The special case for expanding a candidate operator definition {\tt D}
requires a fresh variable for each instance, ensured by Prolog's built-in {\tt copy\_term}.
\begin{code}
expand_defs(D,#(X),R) :-!,copy_term(D,(#(X):-T)),expand_defs(D,T,R).
\end{code}
Other operators are traversed generically by using Prolog's ``\verb~=..~'' built-in and by
recursing with {\tt expand\_def\_list} on their arguments.
\begin{code}
expand_defs(D,A,B) :-
  A=..[F|Xs],
  expand_def_list(D,Xs,Ys),
  B=..[F|Ys].

expand_def_list(_,[],[]).
expand_def_list(D,[X|Xs],[Y|Ys]) :-
  expand_defs(D,X,Y),
  expand_def_list(D,Xs,Ys).
\end{code}
The predicate {\tt prove\_with\_def} refines our {\bf G4ip} prover by first
expanding the definitions extending {\bf IPC} with a given candidate
modality.
\begin{code}
prove_with_def(Def,T0) :-expand_defs(Def,T0,T1),prove_in_ipc(T1,[]).
\end{code}
The definition synthesizer will filter the candidate definitions provided by
{\tt genDef} such that the predicate {\tt prove\_with\_def}  succeeds on all theorems and fails on all
non-theorems, provided as names of the facts of arity 1 containing them. 
\begin{code}
def_synth(M,D):-def_synth(M,iel_th,iel_nth,D).

def_synth(M,Th,NTh,D):-
  genDef(M,D),
  forall(call(Th,T),prove_with_def(D,T)),
  forall(call(NTh,NT), \+prove_with_def(D,NT)).
\end{code}
Note that the generator first builds smaller formulas and then larger ones up the specified maximum size.
\BX Candidate definitions up to size 2
\begin{codex}
?- forall(genDef(2,Def),println(Def)).
#A :- A
#A :- A -> A
#A :- A -> false
#A :- A -> ?
#A :- false -> A
#A :- ? -> A
#A :- A & A
#A :- A & false
#A :- A & ?
...
#A :- (A -> ?) -> A
...
#A :- (? v A) v ?
#A :- (? v false) v A
#A :- (? v ?) v A
\end{codex}
\EX

\section{Discovering the embedding of {\bf IEL} and Dosen's double negation modality in {\bf IPC }}\label{embeil}

We specify a given logic (e.g., {\bf IEL} or {\bf S4}) by stating  theorems on which the prover extended with the synthetic definition should succeed and non-theorems on which it should fail.

\subsection{The discovery mechanism for {\bf IEL}}

We start with the 4 axioms of Artemov and Protopopescu's {\bf IEL} system:
\begin{code}
iel_th(a -> # a).
iel_th(# (a->b)->(# a-> # b)).
iel_th(# p <-> # # p).
iel_th(# a -> ~ ~ a).
\end{code}
Note that the axioms would be enough to specify the logic, but we also add some theorems when intuitively relevant and/or mentioned in \cite{iel16}.
\begin{code}
iel_th(#   (a & b) <-> (# a & # b)).
iel_th(~ # false).
iel_th(~ (# a & ~ a)).
iel_th(~a -> ~ # a).
iel_th( ~ ~ (# a -> a)).
iel_th(# a & # (a->b) -> # b).
iel_th(* (a & b) <-> (* a & * b)).
iel_th(# a -> * a).
iel_th(# a v # b -> # (a v b) ).
iel_th(* a <-> * * a).
iel_th(a -> *a).
\end{code}
Again, following \cite{iel16}, we add our non-theorems.
\begin{code}
iel_nth(# a -> a).
iel_nth(# (a v b) -> # a v # b).
iel_nth(# a).
iel_nth(~ (# a)).
iel_nth(# false).
iel_nth(# a).
iel_nth(~ (# a)).
iel_nth(* false).
\end{code}
We also define (implicit) facts for supporting the {\em necessitation rule}
that states that the operator ``\verb~#~'' applied to proven theorems
or axioms generates new theorems.
\begin{code}
iel_nec_th(T):-iel_th(T).
iel_nec_th(# T):-iel_th(T).
\end{code}
Finally, we obtain the discovery algorithm for {\bf IEL} formula definitions
and for {\bf IEL} extended with the necessitation rule.
\begin{code}
iel_discover:-
  backtrack_over((def_synth(2,iel_th,iel_nth,D),println(D))).
 
iel_nec_discover:-
  backtrack_over((def_synth(2,iel_nec_th,iel_nth,D),println(D))).
  
backtrack_over(Goal):-call(Goal),fail;true. 

println(T):-numbervars(T,0,_),writeln(T). 
\end{code}

We run {\tt iel\_discover}, ready to see the surviving definition candidates.
\BX
Definition discovery without the necessitation rule.
\begin{codex}
?- iel_discover.
#A:-(A->false)->A
#A:-(A->false)->false
#A:-(A-> ?)->A
true.
\end{codex}
\EX

\BX Definition discovery with the necessitation rule.
\begin{codex}
?- iel_nec_discover.
#A:-(A->false)->A
#A:-(A->false)->false
#A:-(A-> ?)->A
true.
\end{codex}
\EX
Unsurprisingly, the results are the same, as a consequence of \verb~A -> #A~.
\begin{codeh}
iel_test:-
   Def=(#A:-(A -> ?) -> A),
   println('theorems'),
   backtrack_over((
     iel_th(T),
     (prove_with_def(Def,T)->Mes=' expected to be PROVEN';Mes='failed'),
     println(T : Mes)
   )),nl,
   println('necessitation rule: if proven A then #A'),
   backtrack_over((
     iel_th(T0),T= #T0,
     (prove_with_def(Def,T)->Mes=' expected to be PROVEN';Mes='failed'),
     println(T : Mes) 
   )),nl,
   println('non-theorems'),
   backtrack_over((
     iel_nth(T),
     (\+ prove_with_def(Def,T)->Mes=' expected to FAIL';Mes='wrongly proven'),
     println(T:Mes)
   )),nl.
\end{codeh}

Clearly, the  formula \verb~#A:-(A->false)->A~ is not interesting as it would define
knowing something as a  contradiction that implies itself.

This brings us to the second definition formula candidate.

\subsection{Eliminating Dosen's double negation modality}
 In [2] double negation in IPC is interpreted as a ``$\Box$'' modality. This corresponds to one of the synthetic definitions \verb~#A :- (A->false)->false~ that is equivalent in {\bf IPC} to
 \verb|#A :- ~~A|. It is argued in \cite{iel16} that it does not make sense as an epistemic modality, mostly because it would entail that all classical theorems are known intuitionistically.

We eliminate it by requiring the collapsing of ``\verb~*~'' into ``\verb~#~'' to be a non-theorem: 
\begin{codex}
iel_nth(* a <-> # a).
\end{codex}
In fact, while {\em known} (\verb|#|) implies {\em knowable} (\verb|~#~ = *|), it is reasonable to think, as in most modal logics, that the inverse implication does not hold.

After that, we have:
\BX
The double negation modality is eliminated, as it collapses \verb~#~ and \verb~*~.

\begin{codex}
?- iel_discover.
#A:-(A -> ?)->A
true.

?- iel_nec_discover.
#A:-(A -> ?)->A
true.
\end{codex}
\EX

\subsection{Knowledge as awareness?}

This leaves us with the \verb~#A :- (A -> ?) -> A~.

Among the consequences of the fact that intuitionistic provability strictly implies classical, is that  there's plenty of room left between {\tt p} and \verb|~~p|, where both
\verb~#~ and \verb~*~ find their place, given that the following implication chain holds.
\begin{codex}
p -> #p -> *p -> ~~p 
\end{codex}

Let us now find an (arguably) intuitive meaning for the ``\verb~?~'' constant
in the definition.
The interpretation of knowledge as awareness about truth goes back to
\cite{awareness85}.
Our final definition of intuitionistic epistemic modality as 
``\verb~#A :- (A -> ?) -> A~'' suggests interpreting ``\verb~?~'' as 
awareness of an agent entailed by (a proof of) \verb~A~.
With this in mind, one obtains an embedding  of {\bf IEL} in {\bf IPC} via the extension
\[
\bk A ~\equiv~ (A \imp {\bf eureka}) \imp  A
\]
 where {\bf eureka} is a new symbol not occurring in the language\footnote{
 Not totally accidentally named, given the way Archimedes expressed his sudden {\em awareness} about the volume of water displaced by his immersed body.
 }.
 
In line with the Brouwer-Heyting-Kolmogorov ({\bf BHK}) interpretation of intuitionistic proof, we may say that an agent {\em knows {\tt A} {\bf iff}  {\tt A} is validated by a proof of {\tt A} that induces awareness of the agent about it}.

Thus knowledge of an agent, in this sense, collects facts that are proven constructively in a way that is ``understood'' by the agent.
The consequence
\[
\bk A \imp \neg \neg A
\]
would then simply say that intuitionistic truths, that the agent is aware of, are also
 classically valid.

Thus, we can define our prover for {\bf IEL} as follows.
\begin{code}
iel_prove(P):-prove_with_def((#A :- (A -> eureka) -> A),P).
\end{code}

Interestingly, if one allows {\tt eureka} to occur in the formulas of the 
language given as input to the prover, then it becomes (the unique) value for which
we have equivalence between being known and having a proof. 
\begin{codex}
?- iel_prove(#eureka <-> eureka).
true .
\end{codex}
Similarly, it would also follow that
\begin{codex}
?- iel_prove(*eureka <-> ~ ~ eureka).
true.
\end{codex}
Thus, one would need to forbid accepting it as part of the prover's language to
closely follow the intended semantics of {\bf IEL}.

\subsection{Discussion}\label{disc}
As the {\bf IPC} fragment with two variables, implication and negation has
exactly {\bf 518} equivalence classes of formulas \cite{db75,deJong}, one would
expect the construction deriving ``\verb~*~'' from ``\verb~#~'' to reach a fixpoint.
We can use our prover to find out when that happens.
\begin{codex}
?- iel_prove(#p <-> ~ # (~p)).
false.
iel_prove(*p <-> ~(*(~p))).
true.
\end{codex}
Thus the fixpoint of the construction is ``\verb~*~'' that we have interpreted
as meaning that a proposition is {\em knowable}. 
Therefore, the equivalence reads reasonably that something is knowable
if and only if its negation is not knowable.
Note also that
\begin{codex}
?- iel_prove(~(*(~p)) -> #p).
false.
\end{codex}
by contrast to the equivalence  $\Box p \equiv \neg \Diamond \neg p$
usual in classical modal logics.

\section{Discovering an embedding of {\bf S4} without the necessitation rule}\label{s4}

The fact that both {\bf IPC} and {\bf S4} are known to be PSPACE-complete
\cite{statman79} means that polynomial-time translations exist between them.

In fact, G\"odel's translation from {\bf IPC} to {\bf S4} (by prefixing each subformula with the
$\Box$ operator) shows that the embedding of {\bf IPC} into {\bf S4} can be achieved
quite easily, by using purely syntactic means.
However, the (very) few papers attempting the inverse translation \cite{egly07,goreS4IPC}
   rely on methods often involving intricate semantic constructions.

We will use our definition generator to identify the problem that precludes
a simple embedding of {\bf S4} into {\bf IPC}.

We start with the axioms of {\bf S4}.
\begin{code}
s4_th(# a -> a).
s4_th(# (a->b) -> (# a -> # b)). 
s4_th(# a -> # # a).
\end{code}
We add a few theorems.
\begin{code}
s4_th(* * a <-> * a).
s4_th(a -> * a).
s4_th(# a -> * a).
s4_th(# a v # b -> # (a v b)). 
s4_th(# (a v b) -> # a v # b). 
\end{code}
We add some non-theorems that ensure additional filtering.
\begin{code}
s4_nth(# a).
s4_nth(~ (# a)).
s4_nth(# false).
s4_nth(* false).
s4_nth(* a -> # * a). % true only in S5
s4_nth(a -> # a).
s4_nth(* a -> a).
s4_nth(# a <-> ?).
s4_nth(* a <-> ?).
\end{code}
Like in the case of {\bf IEL} we define implicit facts
stating that the necessitation rule holds.
\begin{code}
s4_nec_th(T):-s4_th(T).
s4_nec_th(# T):-s4_th(T).
\end{code}
Finally we implement the definition discovery predicates and run them.
\begin{code}
s4_discover:-
  backtrack_over((def_synth(2,s4_th,s4_nth,D),println(D))).   
  
s4_nec_discover:-
  backtrack_over((def_synth(2,s4_nec_th,s4_nth,D),println(D))). 
\end{code}

\BX
The necessitation rule eliminates all simple embeddings of {\bf S4} into {\bf IPC}, while
a lot of definition formulas  pass without it.
\begin{codex}
?- s4_discover.
#A :- A & ?
#A :- ? & A
#A :- A & (A-> ?)
#A :- A & (? -> false)
...
true.

?- s4_nec_discover.
true.
\end{codex}
\EX
Among the definitions succeding without passing the necessity rule test, one might want to pick \verb~#A :- ? & A~ as an approximation of the {\bf S4} 
``$\Box$'' operator.  In this case ``\verb~?~'' would simply state that ``the IPC prover is sound and complete''. Still, given the failure of the necessitation rule, the resulting logic is missing
a key aspect of the intended meaning of {\bf S4}-provability.

\section{Related work}\label{rel}

Program synthesis techniques have been around in logic programming 
with the advent of Inductive Logic Programming \cite{ilp91}, but
the idea of learning Prolog programs 
from positive and negative examples goes back to \cite{shapiro81}.
Our definition synthesizer fits in this paradigm, with focus
on the use of a theorem prover of a decidable logic ({\bf IPC})
filtering formulas provided by a definition generator through
theorems as positive examples and non-theorems as negative examples.
The idea to use the new constant ``\verb~?~'' in our synthesizer is inspired
by proofs that some fragments of {\bf IPC} reduced to two variables
have a (small) finite number of equivalence classes \cite{db75,deJong} as
well as by the introduction of new variables, in work on polynomial
embeddings of {\bf S4} into {\bf IPC} \cite{egly07,goreS4IPC}.

We refer to \cite{iel16} for a thorough discussion of the merits of {\bf IEL} compared to  epistemic logics following closely classical modal logic, but the central idea
about using intuitionistic logic is that of {\em belief and knowledge as the product of verification}. Our embedding of {\bf IEL} in {\bf IPC} can be seen as
a simplified view of this process through a generic ``awareness of an agent'' concept
in line with \cite{awareness85}.

In \cite{gelfond_intro} the concept of {\em epistemic specifications} is introduced that
support expressing knowledge and belief in an Answer Set Programming framework.
Interestingly, refinements of this work like
 \cite{gelfond_new} and \cite{shen16} discuss 
difficulties related to expressing an assumption like 
$p \imp \bk p$ in terms of {\bf ASP-based} epistemic operators.

Equilibrium logic \cite{pearce97} gives a semantics to Answer Set programs by
extending the 3-valued intermediate logic of here-and-there {\bf HT} with Nelson's constructive
strong negation. 
In \cite{Kracht1998} a 5-valued truth-table semantics for equilibrium logic is given.
In \cite{cerro15} (and several other papers)
 epistemic extensions of equilibrium logic \cite{pearce97} are proposed,
 in which $\bk p \imp p$.
 By contrast to ``alethic inspired'' epistemic logics postulating $\bk p \imp p$
we closely follow  the $p \imp \bk p$ view on which \cite{iel16} is centered.

While we have eliminated Dosen's double negation modality \cite{dosen_dneg}
as an epistemic operator $\bk p \equiv \neg \neg p$, it is significant that it came out as the only
other meaningful candidate produced by our definition synthesizer.

This suggests that it might be worth investigating further how  a similar definition
discovery mechanism as the one we have used for {\bf IEL} and {\bf S4}
would work for logics with multiple negation operators like equilibrium logic.

Besides the  $\bk p \imp p$ vs. $p \imp \bk p$ problem
a more general  question is 
the choice of the logic supporting the epistemic operators,
among logics with finite truth-value models
(e.g., classical logic or equilibrium logic)
or, at the limit, intuitionistic logic itself, with no such models.
Arguably, this could be application dependent, 
as epistemic operators built on top of IPC are likely
to fit better the landscape with intricate nuances of a richer set
of epistemic and doxastic operators, while  such operators built
on top of finite-valued intermediate logics would benefit
from simpler decision procedures and faster evaluation mechanisms.

\section{Conclusions}\label{concl}

We have devised a general mechanism for synthesizing definitions that extend a given logic system endowed with a theorem prover. The set of theorems on which the extended prover should succeed and the set of non-theorems on which it should fail can be seen as a declarative specification of the extended system. Success of the approach on embedding the  {\bf IEL} system in {\bf IPC} and failure on trying to embed {\bf S4} has revealed the individual role of the axioms, theorems and rules that specify a given logic system. Given its generality,
our definition generation technique can be applied also to
epistemic or modal logic axiom systems to find out if they have
interesting embeddings in {\bf ASP} and  superintuitionistic logics
for which high quality solvers or theorem provers exist.

\section*{Acknowledgement} 
This research has been supported by NSF grant \verb~1423324~. We thank the anonymous reviewers of the {\bf EELP'2019} workshop for their careful reading, constructive suggestions and comments.

\bibliographystyle{splncs}
\bibliography{theory,tarau,proglang,biblio}

\end{document}